\documentclass[multphys,vecphys]{svmult}

\usepackage{makeidx}         
\usepackage{graphicx}        
\usepackage{multicol}        
\usepackage[bottom]{footmisc}
\RequirePackage{natbib}

\makeindex             

\newcommand{\apj}{ApJ}

\newcommand{\apjl}{ApJL}

\begin{document}

\title*{Beyond the Cool Core: The Formation of Cool Core Galaxy Clusters}
\author{Jack~O.~Burns\inst{1}, Eric~J.~Hallman\inst{1}, Brennan Gantner\inst{1},  Patrick M. Motl\inst{2}, and Michael L. Norman\inst{3} }
\authorrunning{Burns, Hallman, Gantner, Motl, Norman}
\institute{Center for Astrophysics and Space Astronomy, Department of Astrophysical \& Planetary Science, University of Colorado, Boulder, CO 80309 \and
Department of Physics and Astronomy,
Louisiana State University, Baton Rouge, LA 70803 \and
Center for Astrophysics and Space Sciences, University of
California-San Diego, 9500 Gilman Drive, La Jolla, CA 92093}

\maketitle
\section{Abstract}
\label{sec:1}
Why do some clusters have cool cores while others do not?  In this
paper, cosmological simulations,
including radiative cooling and heating, are used to examine the
formation and evolution of cool core (CC) and non-cool core (NCC)
clusters. Numerical CC
clusters at $z=0$ accreted mass more slowly over time and grew
enhanced cool cores via hierarchical mergers; when late major mergers
occurred, the CC's survived the collisions.  By contrast, NCC
clusters of similar mass experienced major mergers early in their
evolution that destroyed embryonic cool cores and produced conditions
that prevent CC re-formation.  We discuss observational consequences.
\section{Introduction}
\label{sec:2}
 Recently, 49\% of clusters were identified as having cool cores in a
flux-limited sample, HIFLUGCS, based upon {\it ROSAT} and {\it ASCA}
observations \cite{chen}. The
earliest and simplest model assumed these clusters to be spherical, isolated
systems where ``cooling flows'' formed; as radiating gas loses
pressure support, cooling gas flows inwards to higher density values
which further accelerates the cooling rate (e.g., \cite{fab02}). 
However, the predicted end-products of this mass infall (e.g., star
formation, HI, CO) have not been observed and the central temperatures
indicate that the gas at the cores has cooled only moderately
(generally only 30-40\% of the virial temperature \cite{dv04}).  Furthermore, this model failed to consider
the important effects of mergers and on-going mass accretion from the
surrounding supercluster environment. 

We have performed numerical simulations of the formation and evolution
of clusters in a cosmological context using the adaptive mesh
refinement N-body/hydro code {\it Enzo}, aimed at further
understanding cool cores \cite{motl04}.  We
find that most mergers are oblique with halos spiraling into the
cluster center and gently bequeathing cool gas; this allows the cool
core to strengthen over time. Thus, in this model, cool cores
themselves grow hierarchically via the merger/accretion process.
Also, a realistic model of cool cores must involve heating, as well as
cooling, that ``softens'' the cores (i.e., reduces density and
increases temperature) potentially by star formation or by AGNs.  This
softening would increase the susceptibility of disruption of the cool
core by subsequent mergers. 
 
In this paper, we present new enhanced-physics {\it Enzo} cosmology
simulations that include radiative cooling, star formation (i.e., a
mass sink for cold gas), and heating (modeled via kinetic energy input
from Type II supernovae).  In addition to the more realistic baryonic
physics, these simulations are superior to previous numerical models
(e.g., \cite{motl,kravtsov}) due to their bigger
volumes and larger samples of clusters.  With hundreds of clusters of
mass $>10^{14} M_{\odot}$, we have the capability for the first time
to examine evolutionary effects in these numerical clusters with
adequate statistics.   
\section{The Formation of CC and NCC Clusters}
\label{sec:3}
Enzo is an adaptive mesh refinement cosmological simulation code that
couples an N-body particle-mesh (PM) solver with an Eulerian
block-structured AMR method for ideal gas dynamics. Enzo has physics
modules which model radiative
cooling of both primordial and metal-enriched gases, and prescriptions for star
formation and feedback \cite{enzo}. 

Our AMR simulations have produced a sample of 1494 numerical clusters
with $0<z<1$ and $M_{virial}>10^{14} M_{\odot}$.  Only $\approx$10\%
of this sample of numerical clusters have cool cores.  This fraction
is smaller than observed.   We believe that our specific recipe for
star formation and heating has over-softened the cores making them too
vulnerable to disruption.  This will be addressed in subsequent
simulations. 

In Figure 1, we show examples of the evolution of a CC and an NCC
cluster which represent well the general scenarios for how such
clusters form.  This figure illustrates the different histories of two
clusters which have the same final mass at $z=0$, $M_{virial}=5 \times
10^{14} M_{\odot}$.   

\begin{figure}
\centering
\includegraphics[width=5cm]{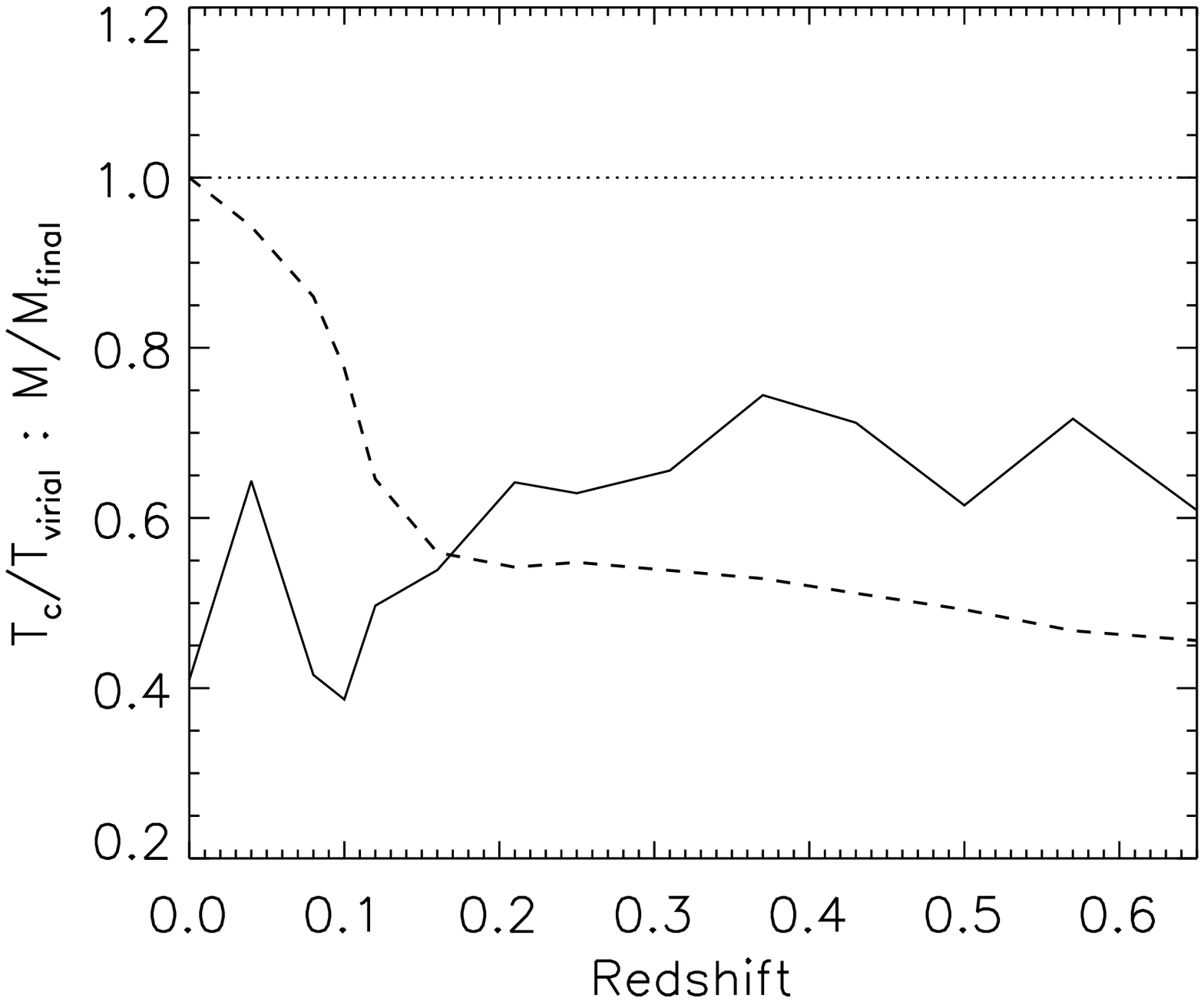}
\includegraphics[width=5cm]{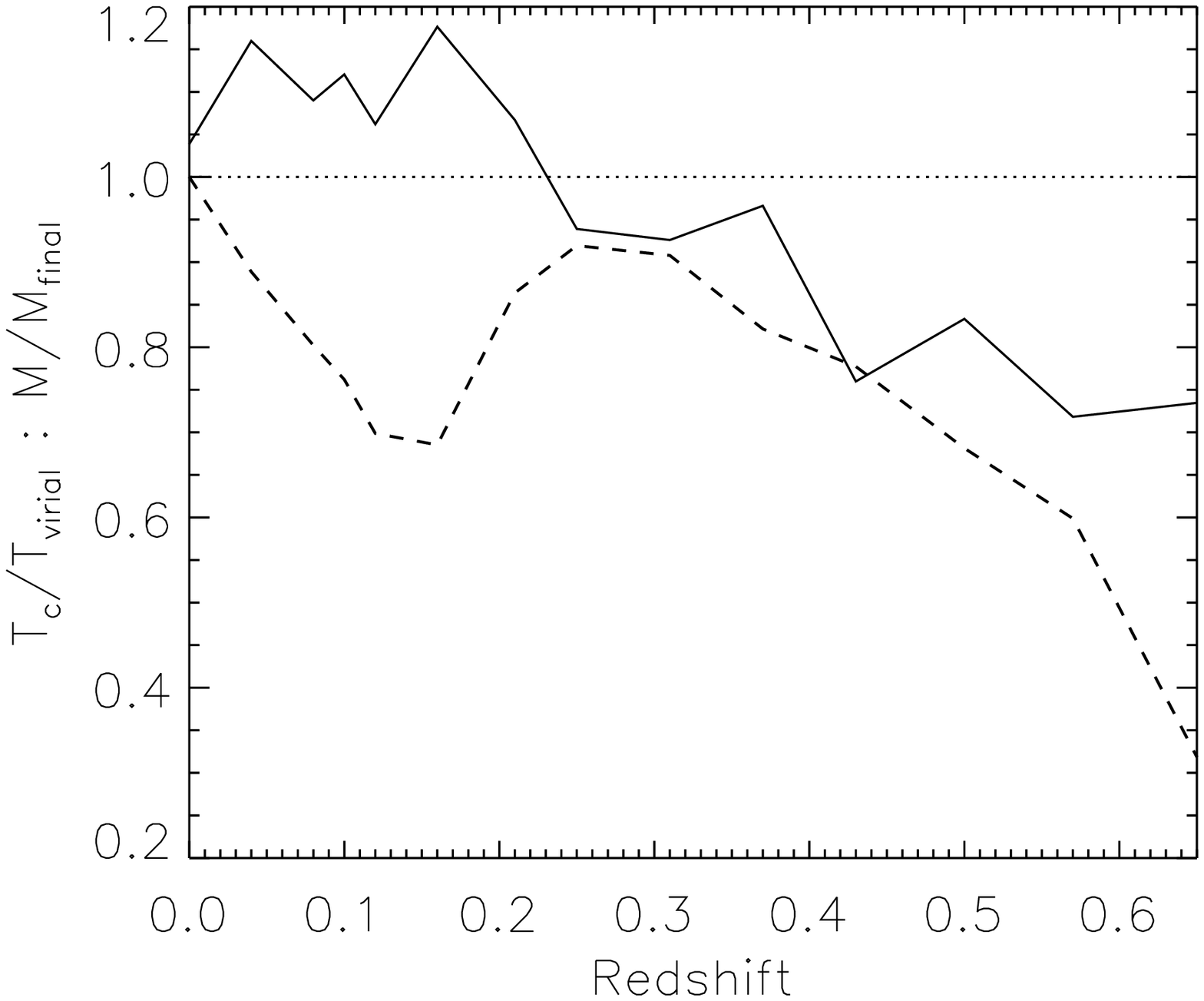}
\caption{The time history of the evolution of a CC (left) and an NCC
  (right) numerical cluster.  Dashed lines indicate the total mass in units of the
  virial mass normalized to 1 at z=0.  Solid line is the central
  temperature normalized by the virial temperature.  Note that the dip
  in mass at $z \approx 0.15$ in the NCC cluster is an artifact of how
  we measure the cluster mass within the virial radius; it is caused
  by the passage of an infalling subcluster through the core and back
  out beyond $r_{virial}$ with a final acquisition into the cluster by
  z=0.} 
\label{fig:1}
\end{figure}
Many clusters start off with cool cores
early in their history when initial conditions produce central
densities and temperatures that allow the gas to radiatively cool.
Depending upon the sequence and magnitude of cluster collisions, cool
cores will be destroyed in some clusters and preserved in others. NCC clusters often
undergo major mergers early in their history.  From z=0.65 to z=0.5,
the mass of the NCC cluster in Fig. 1 increased by 110\%.  Such
mergers destroy the cool cores, leaving behind hotter, thermalized,
moderately dense cores. Subsequently, cool halos infalling into these
NCC clusters do not survive passage through the central parts of the
clusters and the central conditions do not allow cool cores to
re-establish.  NCC clusters continue to experience minor mergers as
they continue to slowly evolve (typically mass increases 10-20\% over
Gyr time frames). We suggest that such an early major merger produced
the complex characteristics observed today for the Coma cluster. 

On the other hand, Figure 1 suggests that CC clusters evolve
differently.  {\it Many CC clusters do not experience major mergers
  early in their history but rather grow slowly such that the cool
  cores increase in mass and stability}.  The CC cluster at z=0.65
began with a higher mass than the NCC at the same redshift. From
z=0.65 to z=0.1, mergers increased the mass of the CC cluster by only
$\approx$20\%.  At $z \approx 0.1$, substantial multiple mergers
occurred increasing the mass by $\approx$50\% by z=0.  But by this
epoch, the cool core is so well established that it survives ever
larger impacts.  In fact, the cool gas from the infalling halos near
the present epoch merges with the cool core such that it continues to grow.  
\section{Consequences of Evolutionary Differences in CC and NCC Clusters}
\label{sec:4}
The large-scale X-ray surface brightness profiles ($S_X$) of our simulated NCC clusters have $S_X$ distributions which fit
very well to $\beta$-models ($S_X \propto [1+(r/r_c)^2]^{1/2
  -3\beta}$) out to the cluster virial radius as we recently described \cite{hall06}. This is not true for CC clusters in our samples.  As
detailed in \cite{hall06}, the $\beta$-model does a poor job of extrapolating
the surface brightness from $r_{500}$ to $r_{200}$ (subscripts refer
to overdensities relative to critical density; $r_{200} \approx
r_{virial}$) for  CC clusters.  The CC cluster profile has a typically
$\approx$35\% steeper slope than the $\beta$-model in the outer part
of the cluster.  We view this as a further signature of recent merger
activity.  The result is a bias or overestimate of the density
profile, leading to a significant overestimate in mass. 
\vspace{-4mm}
\begin{figure}
\centering
\includegraphics[width=7cm]{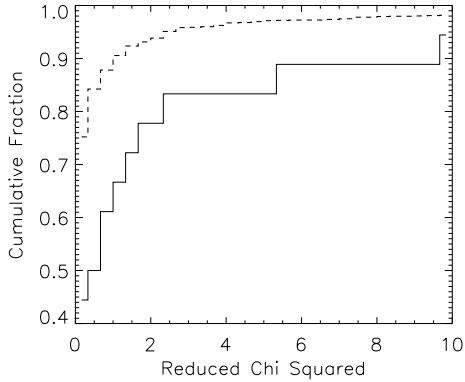}
\vspace{-3mm}
\caption{ Histograms of the quality of fit, as measured by the reduced
  $\chi^2$, for $\beta$-model fits to the X-ray surface brightness
  profiles for $\approx$800 of our numerical clusters with $M>5 \times
  10^{14} M_{\odot}$.  Dotted line is NCC clusters and solid line is
  CC clusters.}   
\label{fig:2}
\end{figure}

These results are typical for CC clusters in our sample. We have fit
$\beta$-models to the synthetic X-ray surface brightness profiles for
all the numerical clusters with $0<z<0.5$ and $M > 5 \times 10^{14}
\mathrm{M_\odot}$ ($\approx$800 clusters).  The fits excluded the core regions with $r<100~h^{-1}$
kpc and extended out to $r_{500}$ to mimic what is typically available
for X-ray observations.  We then extrapolated the fit out to $r_{200}$
and calculated the reduced ${\chi}^2$ between the $\beta$-model and
the actual profile for the numerical clusters.  A histogram of the
cumlative fraction of the ${\chi}^2$ distribution for NCC and CC
clusters is presented in Figure 2.  There is a significant difference
between the quality of the $\beta$-model fits between these two types of
clusters.  Nearly 90\% of the NCC clusters have reduced ${\chi}^2 < 1$
whereas only 60\% of CC clusters have values $< 1$.  This results in a
mass overestimate or bias of typically a factor of 3-5 for CC
clusters. Thus, there is a breakdown in the assumptions of a single
$\beta$-model (i.e., dynamical equilibrium as assumed in simple
cooling flow model) even outside the core for CC
clusters. Similar results for a sample of cool core
clusters observed by {\it Chandra} have been found \cite{vikh06}. 

Our numerical CC clusters also generally show a higher incidence of cold substructure
outside the core than do their NCC counterparts when measured relative
to the virial temperatures.  The lower temperature gas is composed of
both compact infalling halos as well as diffuse cool gas.  We predict
that this signature will be apparent in hardness ratio maps.  As shown
in Fig. 3 for two typical cases drawn from our simulations, the
hard-to-soft band ratios (2-8 keV/0.7-2 keV) illustrate the abundance
of cooler gas beyond the cores in CC clusters.  We find that this
prediction agrees well with the hardness ratio distribution of several
Abell clusters observed by {\it Chandra}.  

\begin{figure}
\centering
\includegraphics[width=10cm]{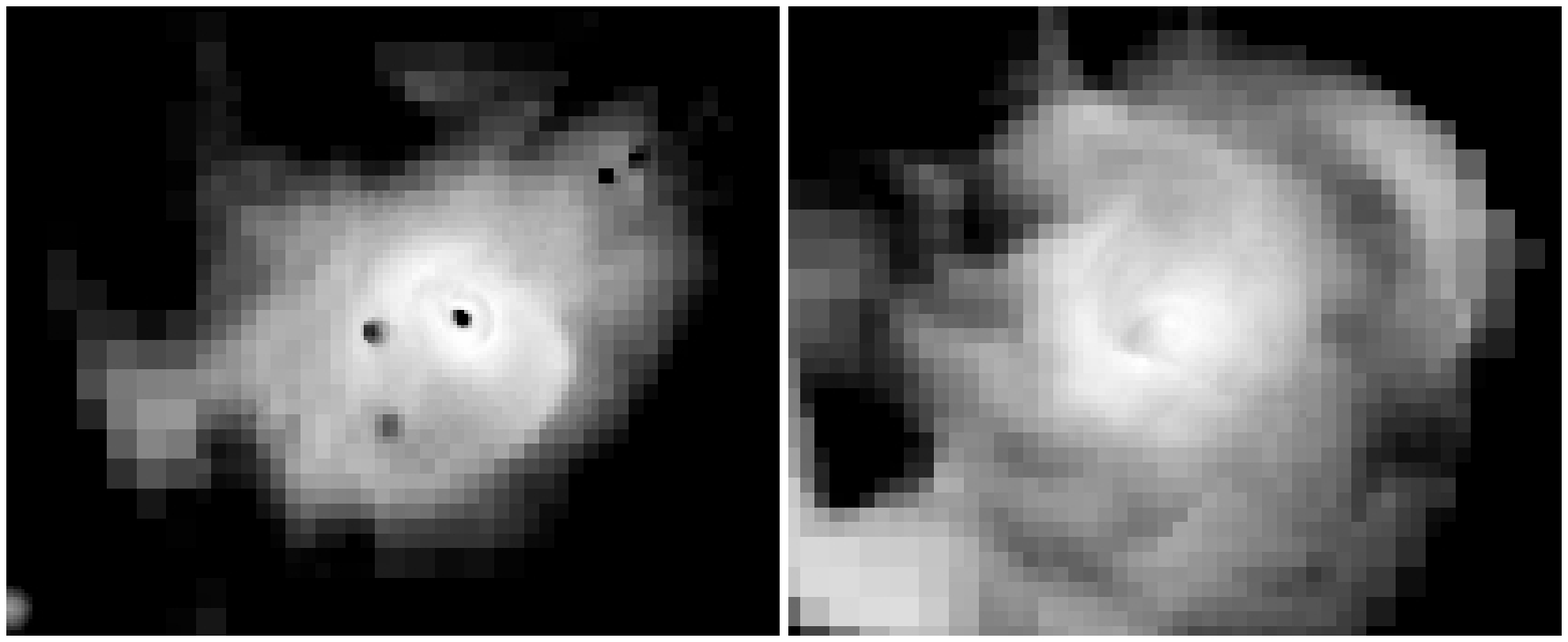}
\includegraphics[width=7cm]{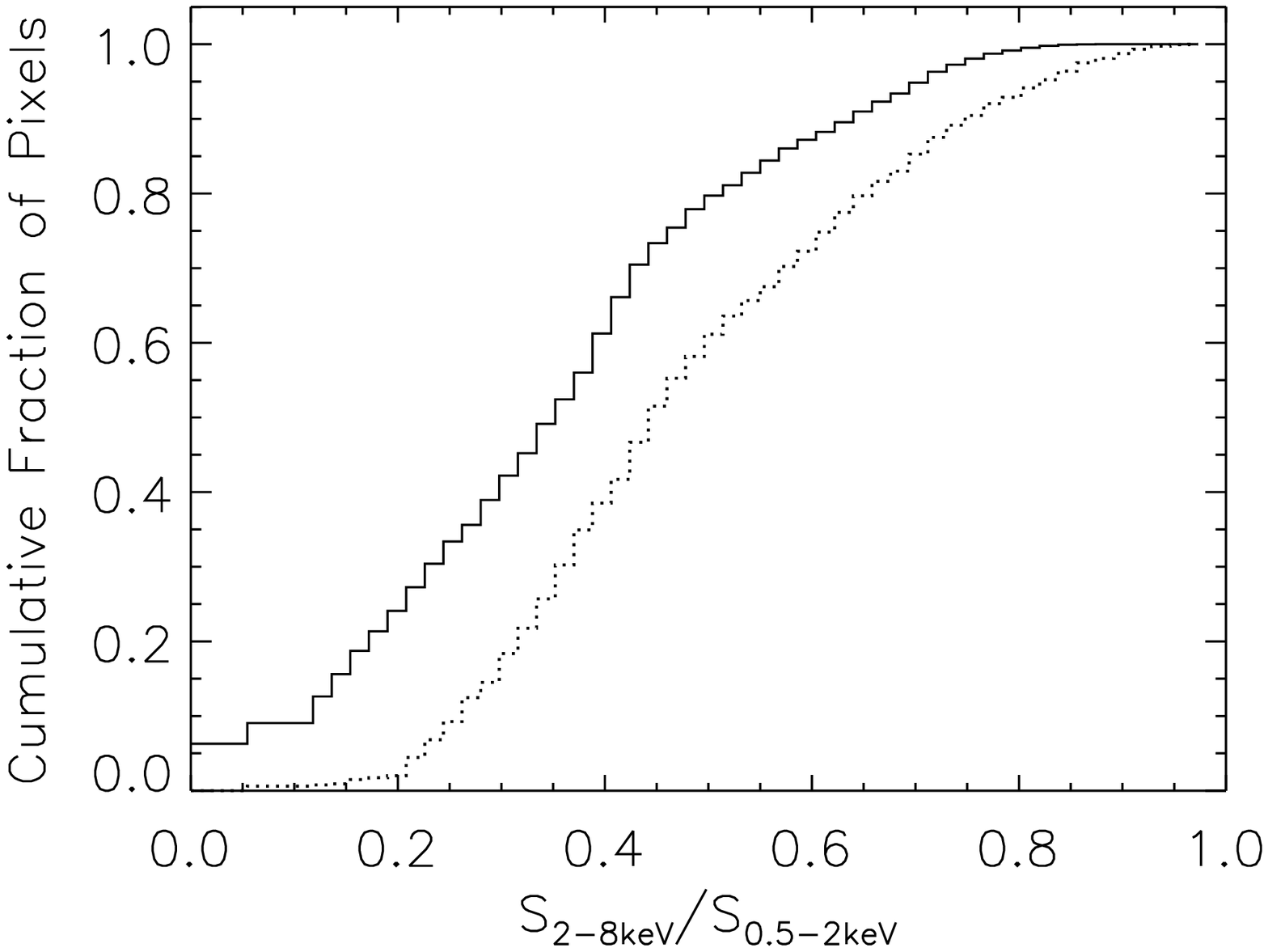}
\caption{Maps (3 $h^{-1}$ Mpc on a side) of the X-ray hardness ratio ($2-8$
keV)/($0.7-2$ keV) for numerical clusters with a cool core (left) and
without a cool core (right).  Dark are smaller (0.2) HR values
(corresponding to cool temperatures) and white are large ($>0.8$) HR
values (hot temperatures).  The clusters have similar masses of
$\approx 5 \times 10^{14} M_{\odot}$.  At the bottom, the histogram is
the distribution of hardness ratios for the two clusters above; dotted is
for the NCC cluster and solid is for the CC cluster. The central 100
kpc of each cluster (cool core region) was excluded in constructing
this histogram.} 
\label{fig:3} 
\end{figure}
\section{Conclusions}
\label{sec:5}
Our cosmological numerical simulations suggest that
\begin{itemize}
\item Non-cool core (NCC) clusters suffer early major mergers when
  embryonic cool cores are destroyed.  Cool core (CC) clusters, on the
  other hand, grow more slowly without early major mergers. 
\item The X-ray surface brightness profiles for NCC clusters are well
  fit by single $\beta$-models whereas the outer emission for CC
  clusters is biased low compared to $\beta$-models.  The resulting
  densities and masses of CC clusters estimated from single
  $\beta$-model extrapolations are biased high by factors of 3-5. 
\item CC clusters have more cool gas beyond the cores than do NCC
  clusters, reflected in hardness ratio maps. 
\end{itemize}
This work was supported in part by grants from the U.S. National Science Foundation (AST-0407368) and NASA (TM3-4008A).

\printindex

\begin{thebibliography}{1}

\bibitem{chen}
Y.~{Chen} and {et al.}
\newblock In H.~{Boehringer}, editor, {\em These proceedings}, 2006.

\bibitem{fab02}
A.~C. {Fabian}.
\newblock {Cooling Flows in Clusters of Galaxies}.
\newblock In {\em Lighthouses of the Universe: The Most Luminous Celestial
  Objects and Their Use for Cosmology: Garching, Germany, 6-10 August 2001, ESO
  ASTROPHYSICS SYMPOSIA. ISBN 3-540-43769-X. Edited by M. Gilfanov, R. Sunyaev,
  and E. Churazov. Springer-Verlag, 2002, p. 24}, pages 24--+, 2002.

\bibitem{dv04}
M.~{Donahue} and G.~M. {Voit}.
\newblock {Cool Gas in Clusters of Galaxies}.
\newblock In J.~S. {Mulchaey}, A.~{Dressler}, and A.~{Oemler}, editors, {\em
  Clusters of Galaxies: Probes of Cosmological Structure and Galaxy Evolution},
  pages 143--+, 2004.

\bibitem{motl04}
P.~M. {Motl}, J.~O. {Burns}, C.~{Loken}, M.~L. {Norman}, and G.~{Bryan}.
\newblock {Formation of Cool Cores in Galaxy Clusters via Hierarchical
  Mergers}.
\newblock {\em \apj}, 606:635--653, May 2004.

\bibitem{motl}
P.~M. {Motl}, E.~J. {Hallman}, J.~O. {Burns}, and M.~L. {Norman}.
\newblock {The Integrated Sunyaev-Zeldovich Effect as a Superior Method for
  Measuring the Mass of Clusters of Galaxies}.
\newblock {\em \apjl}, 623:L63--L66, April 2005.

\bibitem{kravtsov}
A.~V. {Kravtsov}, D.~{Nagai}, and A.~A. {Vikhlinin}.
\newblock {Effects of Cooling and Star Formation on the Baryon Fractions in
  Clusters}.
\newblock {\em \apj}, 625:588--598, June 2005.

\bibitem{enzo}
B.~W. {O'Shea}, G.~{Bryan}, J.~{Bordner}, M.~L. {Norman}, T.~{Abel},
  R.~{Harkness}, and A.~{Kritsuk}.
\newblock {\em {in Adaptive Mesh Refinement: Theory and Applications}}.
\newblock Berlin: Springer, March 2005.

\bibitem{hall06}
E.~J. {Hallman}, P.~M. {Motl}, J.~O. {Burns}, and M.~L. {Norman}.
\newblock {Challenges for Precision Cosmology with X-Ray and Sunyaev-Zeldovich
  Effect Gas Mass Measurements of Galaxy Clusters}.
\newblock {\em \apj}, 648:852--867, September 2006.

\bibitem{vikh06}
A.~{Vikhlinin}, A.~{Kravtsov}, W.~{Forman}, C.~{Jones}, M.~{Markevitch}, S.~S.
  {Murray}, and L.~{Van Speybroeck}.
\newblock {Chandra Sample of Nearby Relaxed Galaxy Clusters: Mass, Gas
  Fraction, and Mass-Temperature Relation}.
\newblock {\em \apj}, 640:691--709, April 2006.

\end{thebibliography}
\end{document}